\documentstyle[preprint,aps]{revtex}
\begin{document}
\title
{Ising transition in dimerized XY quantum spin chain}     
\author{Fei Ye\hspace{1.0cm}Guo-Hui Ding\hspace{1.0cm}Bo-Wei Xu}
\address
{Department of Physics, Shanghai Jiao Tong University, Shanghai 200030,China} \date{\today}
\draft
\maketitle
\begin{abstract}
In the present paper, we proposed a simple spin-1/2 model  which provides
a exactly solvable example to study the Ising criticality with central charge $c=1/2$. 
By mapping it onto the real Majorana fermions, the Ising critical
behavior is explored explicitly, although its bosonized form is not the double frequency
sine-Gordon model. 


\end{abstract}
\newpage

Recently Delfino and Mussardo have shown that in the double frequency
sine-Gordon (DSG)  model there exists a quantum
critical line where the DSG model displays an Ising criticality with central
charge c=1/2, which has stimulated theoretical interest \cite{Delfino,Fabrizio,Bajnok,Ye}. 
This phenomenon is quite general in the various spin chains
and other quantum one-dimensional (1D) systems. By constructing the mappings
between the DSG model and the deformed Ashkin-Teller (DAT) model, Fabrizio
{\it etal} investigated the Ising transition in details \cite{Fabrizio}. They gave the relation
between the original DSG operators and those from the DAT model and identified the degrees
of freedom in the DAT model that become critical. 
They also demonstrated the efficiency of their approach to describe Ising
transition in some physical realizations of the DSG model, such as the
Heisenberg chain in a staggered magnetic field and the 1D Hubbard model with
a staggered potential and accomplished a description of the critical properties 
of the Ising transition. 
In our previous study\cite{Ye}, we analysed the DSG model with
the renormalisation group(RG) technique. Through the one loop $\beta$ equation of the DSG
Hamiltonian, two Ising fixed points and the topology of the RG flows are obtained. Along the 
critical line, the DSG model flows from the central charge c=1 Gaussian fixed point to the
c=1/2 Ising fixed point according to the Zamolodchikov's c-theorem
\cite{Zamolodchikov,Cardy}. 

In this paper we will show the mentioned phenomenon
also occurs in a simpler dimerized XY (DXY) model. Such a model provides us an exactly
solvable example
to show how the Ising criticality with central charge $c=1/2$ occurs, since its low 
energy behavior can be described by two real Majorana fermions, one of which becomes
massless on some critical line.

The Hamiltonian of the DXY model reads
\begin{equation}
H=\sum_n[(1+\gamma)S_n^{x}S_{n+1}^{x}+(1-\gamma)S_n^{y}S_{n+1}^{y}+\delta(-1)^n(S_n^{x}S_{n+1}
^{x}+S_n^{y}S_{n+1}^{y})]\;,\label{hspin}
\end{equation}
where $S_n$ are the spin operators on the lattice sites $n$, $\gamma$ is a
parameter describing the anisotropic XY model, and $\delta$ being the
coupling constant of the dimerized interaction.

It is easy to show that both the anisotropy term  and the dimerization term in the DXY model 
will open a gap $2\gamma$ and $2\delta$, if acting alone, respectively.
However if they coexist, as we shall show, the excitation gap will vanish
by fine tuning of the parameters $\delta$ and $\gamma$ in the DXY model. This implies the
Ising criticality with central charge $c=1/2$, which is revealed obviously by transforming 
the spin operators in DXY model into the real Majorana fields.

Under Jordan-Wigner transformation 
\begin{eqnarray}
&&S^{\dagger}_n=c_n^{\dagger}\exp(i\pi\sum_{l<n}c^\dagger_lc_l)\\\nonumber
&&S^{-}_n=c_n\exp(-i\pi\sum_{l<n}c^\dagger_lc_l)\\\nonumber
&&S^z_n=c^\dagger_nc_n-{1\over 2}\;,
\end{eqnarray}
which maps spins onto spinless fermions, Eq.(\ref{hspin}) becomes
\begin{eqnarray}
H&=&{1\over2}\sum_{n=1}^{N-1}[(1+(-1)^{n}\delta)(c_n^{\dagger}c_{n+1}+h.c.)+
\gamma(c_n^{\dagger}c_{n+1}^{\dagger}+h.c.)]\\\nonumber &&-{1\over 2}[c_N^{\dagger}c_1+
\gamma c_{N}^{\dagger}c_1^{\dagger}+\delta c^{\dagger}_Nc_1+h.c.]\exp(i\pi\sum
_{l=1}^{N}c^{\dagger}_l c_l)\;,\label{hfermion}
\end{eqnarray}
where $c_n$ and $c_n^{\dagger}$ denote the annihilation and creation operators for a fermion on site
$n$, respectively. In Eq.(\ref{hfermion}), $\exp(i\pi\sum_{l=1}^{N}c^{\dagger}_l c_l)$ is conserved and has 
eigenvalues $\pm 1$ which correspond to the cyclic condition and the anticyclic condition,
i.e., $c_1=c_{N+1}$ and $c_1=-c_{N+1}$,
respectively. In the following, we shall focus on the cyclic condition.

For small $\delta$ and $\gamma$, one can express
the fermi operators in terms of the right-moving($R$) and left-moving($L$)
operators near the fermi points
\begin{equation}
c_n=(-i)^n R_n+i^n L_n\;,
\end{equation}
the Hamiltonian density of Eq.(\ref{hfermion}) then can be represented in the continuum
limit $(x=na)$ as
\begin{eqnarray}
{\cal H}&=&-iR^{\dagger}(x)\partial_x R(x)+iL^{\dagger}(x)\partial_xL(x)\\\nonumber
&&-i\gamma[R^{\dagger}(x)L^{\dagger}(x)-L(x)R(x)]+i\delta[R^{\dagger}(x)L(x)
-L^{\dagger}(x)R(x)]\;.\label{hcomplex}
\end{eqnarray}
For our purposes, we introduce two real Majorana fermi fields
$\xi_{R(L)}^{(i)}\;\;\;(i=1,2)$
\begin{equation}
R={1\over\sqrt{2}}(\xi_R^{(1)}+i\xi_R^{(2)})\;\;\;\;\;\;\;\;\;\;\;
L={1\over\sqrt{2}}(\xi_L^{(1)}+i\xi_L^{(2)})\;
\end{equation}
and rewrite Eq.(\ref{hcomplex}) as
\begin{equation}
{\cal H}=-{i\over 2}\sum_{a=1}^2(\xi_R^{(a)}\partial_x\xi_R^{(a)}-\xi_L^{(a)}\partial_x\xi_L^{(a)})
-i(\gamma-\delta)\xi_R^{(1)}\xi_L^{(1)}+i(\gamma+\delta)\xi_R^{(2)}\xi_L^{(2)}\;.\label{hreal}
\end{equation}
Eq.(\ref{hreal}) is the standard form of the Hamiltonian density of two different
Majorana fields with the masses $m_1=\gamma-\delta$ and $m_2=-\gamma-\delta$,
respectively. The Ising criticality is reached when $\gamma=\pm\delta$, one
of the masses vanishes. At this point, the low energy behavior is dominated by the massless
Majorana fermion, which is Ising like, although the bosonized form of the DXY model is not
the DSG model.

Since the Hamiltonian of Eq.(\ref{hfermion}) has a simple quadratic form of fermi operators, it
can be diagonalized 
exactly. From the exact excitation spectrum, one can see that the energy gap vanishes on the critical line $\delta=\pm\gamma$, which confirms the above perturbation theory and indicates
the Ising criticality.

In order to diagonalize the Hamiltonian of Eq.(\ref{hfermion}), we first consider a more general 
quadratic form in fermi operators
\begin{equation}
H=\sum_{i,j}c^{\dagger}_i A_{ij}c_j+{1\over 2}(c^{\dagger}_i B_{ij}c_j^{\dagger}+h.c.)\;.
\label{general}
\end{equation}
The Hermiticity of H requires that {\bf A} be a Hermitian matrix, while the anticommutation 
relationship among the fermi operators require that {\bf B} be an antisymmetric matrix.
Both {\bf A} and {\bf B} are assumed to be real.

Such a Hamiltonian can be reduced to the diagonal form\cite{Lieb} 
\begin{equation}
H=\sum_{k}E_{k}d^{\dagger}_k d_k+{1\over 2}(TrA-\sum_k E_k)\;,
\end{equation}
by a linear transformation
\begin{eqnarray}
&&d_k=\sum_i {\phi_{ki}+\psi_{ki}\over 2}c_i+{\phi_{ki}-\psi_{ki}\over 2}c_i^{\dagger}\;,
\label{trans1}\\
&&d^{\dagger}_k=\sum_i {\phi_{ki}+\psi_{ki}\over 2}c_i^{\dagger}+{\phi_{ki}-\psi_{ki}
\over 2}c_i
\label{trans2}\;,
\end{eqnarray}
where $\phi_{ki}$ and $\psi_{ki}$, considered as real, are to be determined. This transformation should ensure
that the fermi operators $d^\dagger_k$ and $d_k$ obey the anticommutation rules 
$\{d_k,d^\dagger_l\}=\delta_{kl}$ and $\{d_k,d_l\}=0$. By substituting
Eqs.(\ref{trans1},\ref{trans2}) into
\begin{equation}
[d_k,H]-E_kd_k=0\;,
\end{equation}
one can easily find
\begin{eqnarray}
&&\phi_k({\bf A}-{\bf B})=E_k\psi_k\label{phi}\\
&&\psi_k({\bf A}+{\bf B})=E_k\phi_k\;,\label{psi}
\end{eqnarray}
in an matrix notation. Then it is straightforward from Eqs.(\ref{phi},\ref{psi}) to see that
the N-component vectors $\phi_k$ and $\psi_k$ satisfy the following two eigenfunctions
\begin{equation}
\phi_k({\bf A}-{\bf B})({\bf A}+{\bf B})=E^2_k\phi_k\label{eigen1}
\end{equation}
and 
\begin{equation}
\psi_k({\bf A}+{\bf B})({\bf A}-{\bf B})=E^2_k\psi_k\;,\label{eigen2}
\end{equation}
respectively, belonging to the eigenvalue $E_k^2$.
For cyclic matrices {\bf A} and {\bf B}, it is easy to solve the eigenfunctions (\ref{eigen1})
or (\ref{eigen2}). In our case, the relevant matrices are
\begin{equation}
{\bf A}={1\over 2}\left (
\begin{array}{ccccccc}
0 & 1-\delta & 0 & 0 & \cdots & 0 & 1+\delta\\
1-\delta & 0 & 1+\delta & 0 & \cdots & 0 & 0\\
0 & 1+\delta & 0 & 1-\delta & \cdots & 0 & 0\\
0 & 0 & 1-\delta & 0 & \cdots & 0 & 0\\
\vdots&\vdots&\vdots&\vdots&\ddots&\vdots&\vdots\\
0 & 0 & 0 & 0 & \cdots & 0 & 1-\delta\\
1+\delta & 0 & 0 & 0 & \cdots & 1-\delta &0\\
\end{array}
\right )
\end{equation}
and
\begin{equation}
{\bf B}={1\over 2}\left (
\begin{array}{ccccccc}
0&\gamma&0&0&\cdots&0&-\gamma\\
-\gamma&0&\gamma&0&\cdots&0&0\\
0&-\gamma&0&\gamma&\cdots&0&0\\
0&0&-\gamma&0&\cdots&0&0\\
\vdots&\vdots&\vdots&\vdots&\ddots&\vdots&\vdots\\
0&0&0&0&\cdots&0&\gamma\\
\gamma&0&0&0&\cdots&-\gamma&0\\
\end{array}
\right )\;.
\end{equation}
Then, the Hermitian matrix ${\bf (A+B)(A-B)}$ reads
\begin{equation}
{1\over 2}\left (
\begin{array}{ccccccc}
1+(\delta-\gamma)^2&0&{1\over 2}[1-(\delta-\gamma)^2]&0&\cdots&{1\over 2}[1-(\delta-\gamma)^2]&0\\
0&1+(\delta+\gamma)^2&0&{1\over 2}[1-(\delta+\gamma)^2]&\cdots&0&{1\over 2}[1-(\delta+\gamma)^2]\\
{1\over 2}[1-(\delta-\gamma)^2]&0&1+(\delta-\gamma)^2&0&\cdots&0&0\\
0&{1\over 2}[1-(\delta+\gamma)^2]&0&1+(\delta+\gamma)^2&\cdots&0&0\\
\vdots&\vdots&\vdots&\vdots&\ddots&\vdots&\vdots\\
{1\over 2}[1-(\delta-\gamma)^2]&0&0&0&\cdots&1-(\delta-\gamma)^2&0\\
0&{1\over 2}[1-(\delta+\gamma)^2]&0&0&\cdots&0&1-(\delta+\gamma)^2\\
\end{array}
\right ) \;\label{ab},
\end{equation}
which are cyclic with period 2.
To show how to calculate the eigenvalue of the matrix  of
Eq.(\ref{ab}), we consider a general cyclic matrix with period 2, which has the form
\begin{equation}
{\bf M}=\left (
\begin{array}{cccccc}
a_1&a_2&a_3&a_4&\cdots&a_N\\
b_1&b_2&b_3&b_4&\cdots&b_N\\
a_{N-1}&a_N&a_1&a_2&\cdots&a_{N-2}\\
b_{N-1}&b_N&b_1&b_2&\cdots&b_{N-2}\\
\vdots&\vdots&\vdots&\vdots&\ddots&\vdots\\
b_3&b_4&b_5&b_6&\cdots&b_2
\end{array}
\right )\;.
\end{equation}
It has eigenvectors 
\begin{equation}
V_k=(\epsilon,\xi\epsilon^2,\epsilon^3,\xi\epsilon^4,\cdots,\xi\epsilon^N)^T\;\label{vector}
\end{equation} with 
$\epsilon=\exp(ik),k=(2n\pi)/N(n=0,1,2,\cdots,N)$ and parameter $\xi$ to be determined, 
corresponding the eigenvalue $\lambda_k$.
The real $\phi_k$ and $\psi_k$ can be extracted from the real or imaginary part of $V_k$, since
the matrice $\bf (A\pm B)(A\mp B)$ are real in our case.
Substituting Eq.(\ref{vector}) into the eigenfunction 
$({\bf M}-\lambda_k{\bf I})V_k=0$, 
one obtains
\begin{eqnarray}
\lambda_k&=&a_1+\xi a_2\epsilon+\cdots+\xi a_N\epsilon^N\\ \nonumber
&=&(\xi\epsilon)^{-1}(b_1+b_2\xi\epsilon+b_3\epsilon^2+\cdots+b_N\xi\epsilon^{N-1})\;.\label{ddd}
\end{eqnarray}
For the matrix ${\bf (A+B)(A-B)}$ we are interested in, the eigenvalues $E^2_k$ are 
$\cos^2k+(\delta+\gamma)^2\sin^2k$ and $\cos^2k+(\delta-\gamma)^2\sin^2k$.

Consequently, the energy spectrum of the Hamiltonian of Eq.(\ref{hfermion}) are written as
\begin{eqnarray}
&&E_k^{(+)}=\pm\sqrt{\cos^2k+(\delta+\gamma)^2\sin^2k}\;,\label{sa}\\
&&E_k^{(-)}=\pm\sqrt{\cos^2k+(\delta-\gamma)^2\sin^2k}\;.\label{sb}
\end{eqnarray}

Eqs.(\ref{sa},\ref{sb}) indicate that the energy spectrum has  two branches, one with a 
gap $2|\delta+\gamma|$ and 
the other with a gap $2|\delta-\gamma|$, which correspond to the two massive Majorana 
fermions, respectively. The exact solution shows unambiguously that one of the excitation gap 
vanishes when $\delta=\pm\gamma$, where the Ising phase transition occurs.

In summary, in the present paper, we propose a spin-1/2 model DXY chain which provides
a simple example to study the Ising criticality with central charge $c=1/2$. By mapping 
it onto the real Majorana fermions and diagonalizing it exactly, the Ising critical
behavior is explored explicitly, although its bosonized form is not the DSG model. 
It seems that the Ising phase transition is quite a general feature for a variety of spin chain
systems besides those which can be transformed into the DSG model.

The project is supported by the National Natural Science Foundation of China(Grant No.19975031)
and the RFDP(Grant No.199024833).

\end{document}